# Vertical β-$Ga_2O_3$ Schottky Diodes with Deep-Etch Field Termination using Plasma-free Low-pressure CVD-based Ga-assisted Etching

Saleh Ahmed Khan, Ahmed Ibreljic, and Anhar Bhuiyan*
Department of Electrical and Computer Engineering, University of Massachusetts Lowell, MA 01854 USA
*Corresponding author email: anhar_bhuiyan@uml.edu

*Abstract*— **A deep-etch field termination strategy by using a Ga-assisted plasma-free etching technique in a low-pressure chemical vapor deposition (LPCVD) system is demonstrated for β-$Ga_2O_3$ Schottky barrier diodes (SBDs). The thermally activated etching method provides a plasma-free route to form deep mesa terminations while maintaining excellent device integrity. The fabricated diodes exhibit ideal forward conduction with a turn-on voltage of 1.14V, a Schottky barrier height (SBH) of 1.15eV, an ideality factor of 1.20, and a specific on-resistance of 3.72 mΩ·cm², all closely matching the unetched planar devices. Capacitance-voltage analysis further confirms a uniform carrier concentration of $2\times10^{16}$ $cm^{-3}$ and an SBH of 1.23eV, indicating stable electrical characteristics after deep mesa formation. Temperature dependent electrical measurements from 25 to 250°C show stable thermionic-emission transport behavior, with a gradual rise in on-resistance at higher temperatures from phonon-limited mobility. Over this range, the SBH decreases from 1.16 to 1.12eV, while the ideality factor increases from 1.21 to 1.33. The leakage current remains low across the entire temperature range, and the rectification ratio stays above $10^5$ even at 250°C. Under reverse bias, the diodes exhibit an increase in breakdown voltage from 287 to 500V, confirming the effectiveness of geometric field redistribution achieved by the deep-etched mesa. TCAD simulations corroborate these observations, showing suppression of electric field crowding near the anode edge. Together, these results establish Ga-assisted etching as a reliable and plasma-damage-free method for field termination in β-$Ga_2O_3$ devices.**

## I. Introduction

Ultra-wide bandgap β-$Ga_2O_3$ semiconductor has attracted significant attention for next-generation high-voltage power electronics because of its large bandgap (4.5-4.9 eV) and high critical field (8 MV $cm^{-1}$) [1-3]. These intrinsic properties yield a Baliga figure of merit surpassing those of SiC and GaN [4], offering a path toward compact, thermally robust, and energy-efficient power devices. Furthermore, bulk β-$Ga_2O_3$ substrates can be produced using low-cost melt-based growth techniques [5], while high-quality thick epitaxial layers can be realized through vapor-phase growth methods [6-16], enabling vertical device architectures with scalable current and voltage handling [4, 16-22]. However, the practical performance of vertical β-$Ga_2O_3$ Schottky barrier diodes (SBDs) remains constrained by electric-field crowding at the Schottky anode perimeter, which causes premature breakdown at voltages far below the material limit [23-25]. Efficient management of edge electric fields is therefore essential to enhance blocking capability while maintaining low forward conduction loss.

A wide range of termination schemes has been investigated in wide-bandgap technologies, including field plates, junction-termination extensions (JTEs), and guard rings, which have demonstrated strong improvements in GaN [26] and SiC [27] diodes. When applied to β-$Ga_2O_3$, similar approaches have enabled breakdown voltages above 2-3 kV in vertical structures [28-33], but their scalability is hindered by dielectric reliability issues and interface-charge instabilities, which limit termination efficiency [34-36]. To overcome these constraints, geometric termination using deep mesa or trench etching has emerged as an alternative route to redistribute the electric field at the anode edge [21, 22, 37-44]. This approach eliminates the need for dielectric layers and instead relies on physical removal of the laterally depleted region. Deep-etch termination has been highly effective in GaN [45] and SiC [46] power devices and was recently adapted to β-$Ga_2O_3$ [22, 37, 40, 47], revealing that deep mesa etches can suppress the sharp electric-field peaks that form at the anode perimeter and significantly improve field-termination efficiency. Vertical β-$Ga_2O_3$ SBDs incorporating a 4 µm deep plasma-etched mesa exhibited ~3.5-fold enhancement in breakdown voltage without notable degradation in on-state characteristics [22]. However, chlorine-based plasma etching, commonly used for $Ga_2O_3$ trench or mesa fabrication, introduces surface and sidewall damage due to ion bombardment, leading to charge compensation and anisotropic depletion along different crystallographic orientations [48-50]. This challenge motivates the exploration of plasma-free etching techniques that preserve the crystalline integrity and chemical stoichiometry of β-$Ga_2O_3$.

Several such methods have been explored, including wet chemical etching [51], metal-assisted chemical etching (MacEtch) [52, 53], and *in situ* halide or sub-oxide etching within molecular beam epitaxy (MBE) [54], metalorganic chemical vapor deposition (MOCVD) [55-57], or halide vapor phase epitaxy (HVPE) [58] environments. Among them, Ga-assisted low-pressure chemical vapor deposition (LPCVD) etching has recently emerged as a robust plasma-damage-free alternative [59, 60]. The process exploits the thermally driven reaction: 4 Ga + $Ga_2O_3$ → 3 $Ga_2O$ ↑, where volatile $Ga_2O$ sub-oxides desorb under oxygen-deficient conditions, enabling smooth, anisotropic removal of $Ga_2O_3$ without plasma exposure. This etching mechanism is crystallographically selective and preserves the surface stoichiometry and lattice integrity of the material [61], providing a versatile route for scalable and contamination-free device processing. Using this LPCVD-based approach, quasi-vertical β-$Ga_2O_3$ Schottky diodes fabricated entirely through all-LPCVD growth and in-situ Ga-assisted etching demonstrated excellent rectification, low specific on-resistance (8.6 mΩ·cm²), and stable thermionic-emission transport up to 250 °C, confirming the structural and electronic integrity of LPCVD-etched surfaces [61].

Building on these developments, this work introduces a plasma-free LPCVD-based approach for edge field termination in vertical β-$Ga_2O_3$ SBDs. Commercial HVPE-grown β-$Ga_2O_3$ epilayer on native substrates is used, and deep mesa structures are formed by Ga-assisted LPCVD etching to act as the termination geometry. This approach leverages the geometric field-redistribution provided by the mesa while retaining the damage-free characteristics of Ga-assisted etching. The results show that LPCVD etching enhances breakdown capability while preserving stable forward and thermal behavior, indicating that the crystalline and electronic integrity of the β-$Ga_2O_3$ drift layer is maintained.

## II. Experimental Details

Vertical β-$Ga_2O_3$ SBDs were fabricated on (001) β-$Ga_2O_3$ epitaxial wafers grown by halide vapor phase epitaxy (HVPE) (Novel Crystal Technology, Inc.). The epi-structure comprised a ~10.8 µm-thick $n^-$ drift layer (Si-doped, $2 \times 10^{16}$ $cm^{-3}$) grown on a ~625 µm-thick Sn-doped $n^+$ substrate (~$5.4 \times 10^{18}$ $cm^{-3}$). Edge termination was achieved through plasma-free Ga-assisted LPCVD deep mesa etching. A 100 nm-thick $SiO_2$ hard mask was deposited by plasma-enhanced chemical vapor deposition (PECVD) and patterned by optical lithography to expose the edge region surrounding the Schottky anode. Figure 1 shows the complete fabrication flow, including $SiO_2$ mask deposition and patterning, Ga-assisted mesa etching, mask removal, and final metallization steps. The etching was performed in a custom-built horizontal LPCVD system [60, 61]. High-purity metallic Ga (99.99999%) served as the solid etchant source, and high-purity Ar (99.9999%) was used as the carrier and purge gas. The samples were positioned 2 cm downstream from the Ga source, and the process was carried out at 1050 °C and a chamber pressure of 1.2 Torr. The etch duration of 2.5 hours produced a mesa depth of 8.4 µm, corresponding to an average etch rate of 3.36 µm $h^{-1}$. After etching, the $SiO_2$ mask was stripped in buffered oxide etch (BOE). The samples were then rinsed in deionized water and dried in $N_2$. Following mesa formation, the ohmic contact was first formed on the back side of the $n^+$ β-$Ga_2O_3$ substrate by blanket deposition of Ti/Au (30 nm/100 nm) using electron-beam evaporation, followed by rapid thermal annealing (RTA) at 470 °C for 1 min in $N_2$ ambient. Subsequently, the Schottky contact was defined on the top surface by depositing Ni/Au (30 nm/100 nm) through a photolithographically patterned lift-off process. In addition, to evaluate the effect of the 1050 °C LPCVD thermal cycle used during Ga-assisted etching, planar SBDs were fabricated on a separate diced piece from the same wafer without exposure to the high-temperature LPCVD etching step. Electrical characterization was performed on a temperature-controlled probe station. Current density-voltage (J-V) and capacitance-voltage (C-V) measurements were obtained using a Keithley 4200A-SCS parameter analyzer.

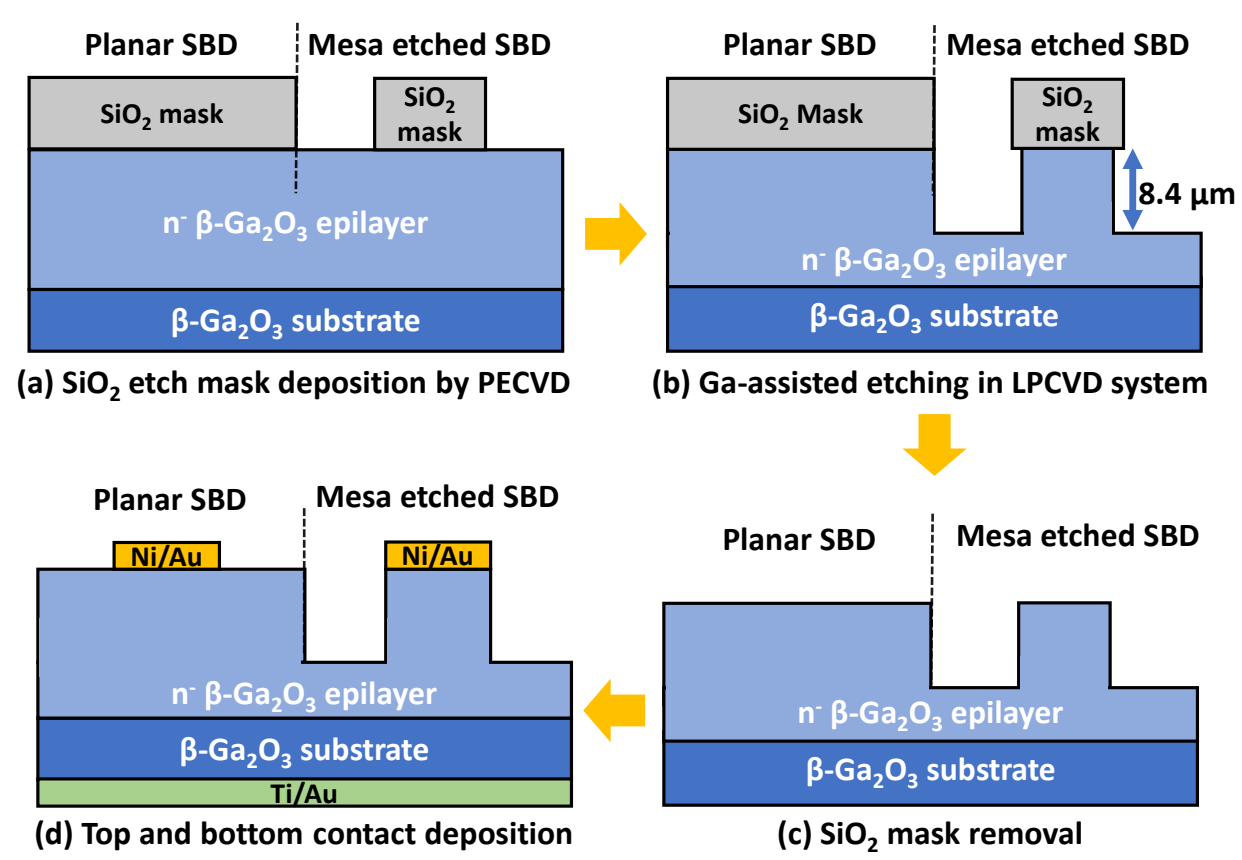

Fig. 1. Fabrication process flow for planar and mesa-terminated β-$Ga_2O_3$ Schottky barrier diodes. (a) A 100 nm thick $SiO_2$ hard mask is deposited by PECVD and patterned by photolithography to define the mesa-etch regions. (b) Ga-assisted LPCVD etching is performed through the $SiO_2$ openings to create a recessed mesa structure with a depth of 8.4 µm in selected regions of the wafer, while adjacent regions remain unetched for planar devices. (c) The $SiO_2$ mask is removed after etching by using BOE. (d) Ti/Au ohmic contacts are formed on the backside followed by rapid thermal annealing at 470 °C for 1 min, after which Ni/Au Schottky contacts are deposited on the front side.

## III. Results and Discussions

Figure 2(a) shows the schematic structure of the fabricated vertical β-$Ga_2O_3$ SBDs with an 8.4 µm recessed mesa formed around the anode perimeter. Cross-sectional SEM analysis, shown in Fig. 2(c), reveals a slightly inclined mesa sidewall. The extracted height profile in Fig. 2(d) indicates a mesa depth of ~8.4 µm with a sidewall angle of 75°, consistent with the crystallographic anisotropy of the LPCVD Ga-assisted etching process. Figure 2(b) compares the forward and reverse J-V characteristics of the planar and LPCVD mesa-terminated β-$Ga_2O_3$ SBDs. Both devices exhibit strong rectification with on/off ratios exceeding $10^8$ and smooth exponential behavior in the forward bias regime, which confirms well behaved thermionic emission transport. The nearly overlapping forward characteristics indicate that the plasma free LPCVD deep etch does not introduce measurable interface states or surface damage. The forward conduction parameters are extracted using the standard thermionic emission (TE) formalism [62, 63] applied to the measured J-V curves,

$$J = J_s\left[exp\left(\frac{qV}{\eta k_0 T}\right) - 1\right] \quad (1)$$

$$J_s = A^* T^2\, exp\left(-\frac{q\varphi_B^{JV}}{k_0 T}\right) \quad (2)$$

$$A^* = \frac{4\pi q m_n^* k_0^2}{h^3} \quad (3)$$

where q is the electric charge, $k_0$ is the Boltzmann constant, and η is the ideality factor, $J_s$ is the reverse saturation current density, $\varphi_B^{JV}$ is the apparent Schottky barrier height (SBH) obtained from the TE model, and $A^*$ is Richardson's constant (41.04 $Acm^{-2}K^{-2}$) [64]. Since the apparent barrier height extracted from Equation (2) depends sensitively on η, a correction is applied to obtain a more accurate barrier height [19, 65], as described by Wagner et al [66].

$$\varphi_B = \left(\varphi_B^{JV} - \frac{\eta - 1}{\eta}\frac{kT}{q} ln\frac{N_C}{N_D}\right)\eta \quad (4)$$

where $N_C$ is the effective conduction band density of states and $N_D$ is the donor concentration. The planar SBD exhibits an ideality factor (η) of 1.19 and a SBH ($q\varphi_B$) of 1.14 eV at room temperature, while the LPCVD mesa etched device shows η = 1.20 and $q\varphi_B$ = 1.15 eV. These values confirm a nearly ideal transport process with negligible barrier inhomogeneity [67-69]. A slight increase in turn-on voltage from 1.05 V to 1.14 V is observed. Importantly, the close similarity of η and $q\varphi_B$ between the planar and etched diodes demonstrates that the LPCVD process preserves the Schottky interface energetics. The specific on resistance $R_{on,sp}$ extracted from the linear forward bias region remains essentially unchanged, increasing only slightly from 3.55 to 3.72 mΩ·cm² after mesa etching. This minimal change indicates that the Ga assisted LPCVD process does not degrade drift layer conductivity or mobility. The values observed here are also lower than those reported for 4 µm deep ICP-RIE-terminated SBDs, where higher $R_{on,sp}$ has been noted after plasma etching of the mesa [22].

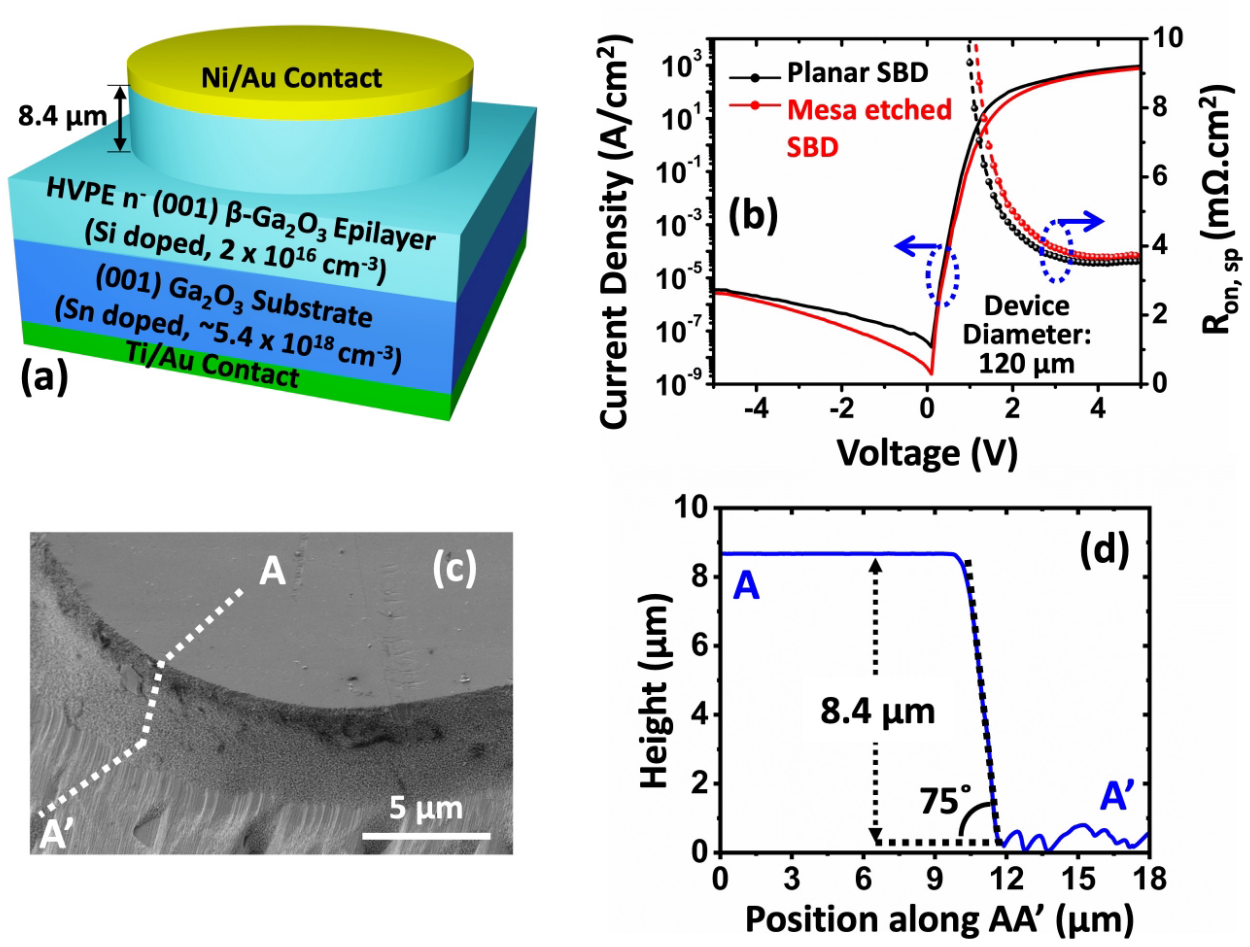

Fig. 2. (a) Schematic illustration of the mesa-terminated β-$Ga_2O_3$ Schottky barrier diode. (b) Comparison of the forward and reverse J-V characteristics of the planar and mesa-etched SBDs. (c) Tilted (45°) cross-sectional SEM image of the etched mesa region, illustrating the sidewall profile. (d) Corresponding height profile extracted along the A-A′ direction, indicating an etch depth of ~8.4 µm and a sidewall inclination of ~75°.

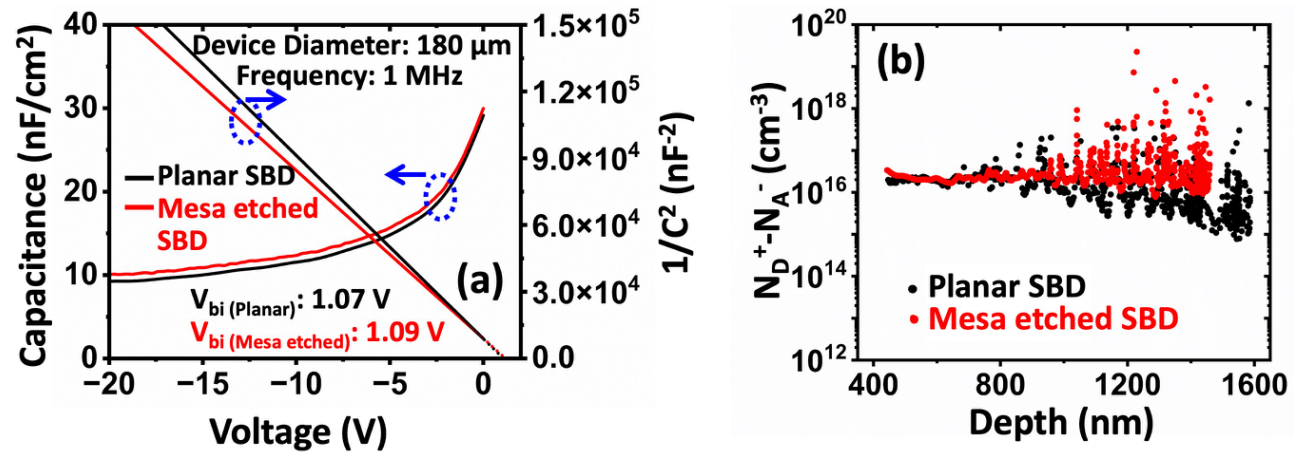

Fig. 3. (a) C-V and corresponding $1/C^2$ - V characteristics of planar and mesa-etched β-$Ga_2O_3$ SBDs. (b) Extracted net donor concentration profiles for both devices, indicating comparable and uniform doping across the measured depletion depth.

The C-V characteristics and extracted carrier profiles of the planar and LPCVD mesa-terminated diodes are shown in Figure 3. Both devices exhibit a nearly linear $1/C^2$-V dependence (Figure 3a), consistent with a uniformly doped n-type drift layer and depletion-region modulation that follows Schottky junction electrostatics. The built-in potential ($V_{bi}$), SBH ($\Phi_B$) and net carrier concentration ($N_d^+$-$N_a^-$) are determined using the following relations [62] for Schottky diodes,

$$N_d^+ - N_a^- = \frac{2}{q\varepsilon_r\varepsilon_0 A^2 (\frac{d\frac{1}{C^2}}{dV})} \tag{5}$$

$$\frac{A^2}{C^2} = \left[\frac{2}{q\varepsilon_r\varepsilon_0 (N_d^+ - N_a^-)}\right]\left(V_{bi} - \frac{kT}{q} - V\right) \tag{6}$$

$$\varphi_B = V_{bi} + \frac{KT}{q} ln\left[\frac{N_C}{N_d^+ - N_a^-}\right] \tag{7}$$

where $\varepsilon_r$ = 10 is the relative permittivity of β-$Ga_2O_3$, $A$ is the device area, $N_C = 5.2\times10^{18}$ cm$^{-3}$ is the density of states in the conduction band. The doping profiles extracted from the C-V data are presented in Figure 3(b). Both the planar and mesa-etched diodes exhibit consistent net donor concentrations of 2 × $10^{16}$ cm$^{-3}$ across the measured depletion depth, suggesting that the high-temperature Ga-assisted LPCVD etching does not noticeably affect the doping profile near the junction. This observation is consistent with prior studies of β-$Ga_2O_3$/$SiO_2$ interfaces, which report chemically abrupt interfaces without clear evidence of significant Si penetration into β-$Ga_2O_3$ after high-temperature processing [70, 71]. The extracted $V_{bi}$ increases slightly from 1.07 V for the planar diode to 1.09 V for the mesa-terminated diode, consistent with the modest increase in turn-on voltage observed in the J-V characteristics. The Schottky barrier heights obtained from C-V analysis are 1.22 eV for the planar diode and 1.23 eV for the mesa-terminated diode, in good agreement with the values extracted from the J-V measurements.

To evaluate the effect of the high-temperature LPCVD processing cycle associated with the Ga-assisted etching procedure (1050 °C), we fabricated two sets of nominally identical planar β-$Ga_2O_3$ SBDs on diced pieces originating from the same HVPE-grown epiwafer. Both samples followed identical device fabrication procedures. The only intentional difference between the two sets was that one sample underwent the 1050 °C LPCVD thermal cycle used for the Ga-assisted etching process, with a patterned $SiO_2$ hard mask present on the surface, whereas the control sample was processed in parallel without exposure to this high-temperature LPCVD step. The corresponding electrical characteristics are shown in Fig. 4. As shown in Fig. 4(a), both devices exhibit nearly identical forward and reverse J-V characteristics, with specific on-resistances of 3.55 and 3.52 mΩ·cm² and ideality factors of 1.19 for both devices. The turn-on voltages (1.05 and 1.02 V) and J-V-extracted barrier heights (1.14 and 1.13 eV) also remain nearly unchanged. Figure 4(b) shows comparable reverse-blocking behavior, with breakdown voltages of 287 V and 302 V for the annealed and unannealed devices, respectively. Likewise, the C-V characteristics in Fig. 4(c) yield nearly identical built-in potentials (1.072 and 1.075 V) and barrier heights (1.215 and 1.218 eV), while the extracted carrier concentration profiles shown in Fig. 4(d) remain at ~$2 \times 10^{16}$ $cm^{-3}$ throughout the measured depletion region. These results indicate that the 1050 °C LPCVD thermal cycle associated with the Ga-assisted etching process do not measurably affect the drift-layer doping, Schottky-interface properties, or carrier transport characteristics.

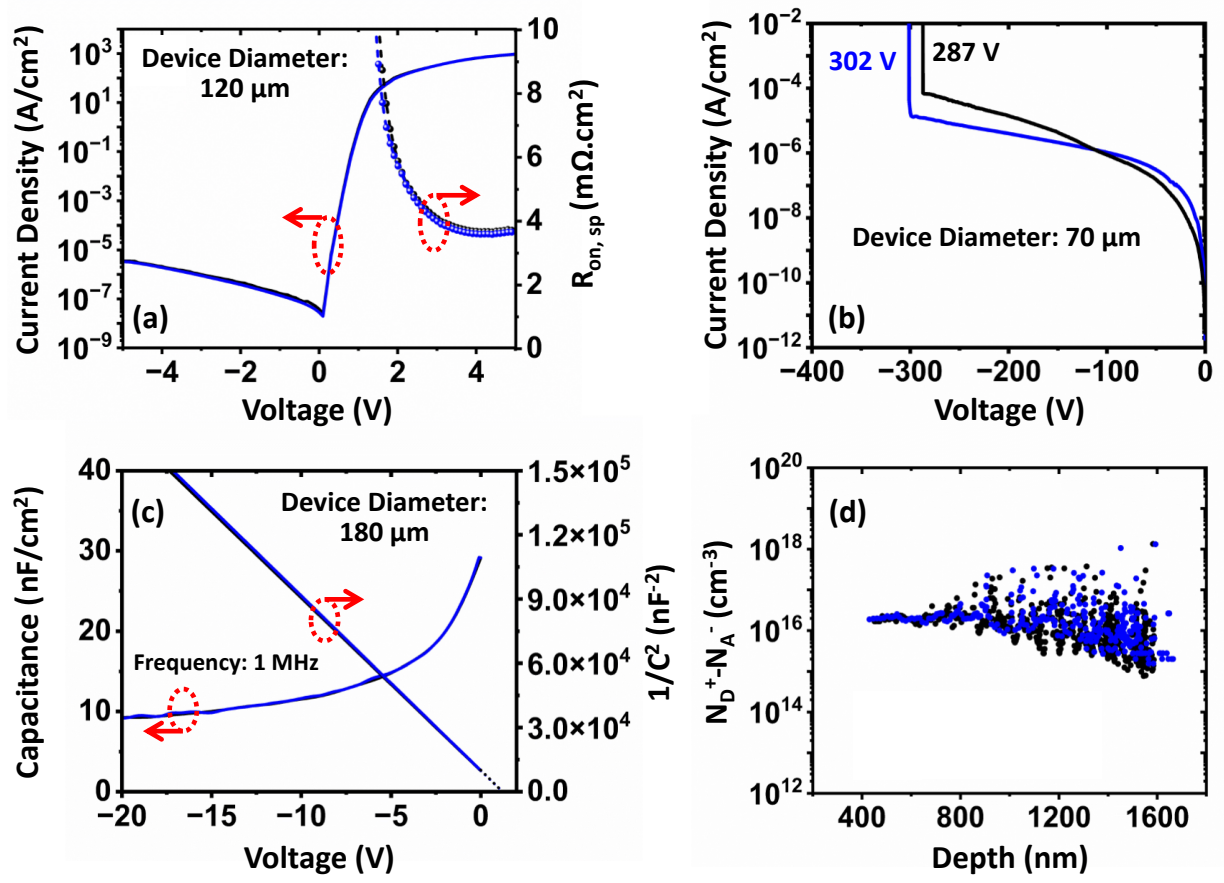

Fig. 4. Electrical comparison of planar β-$Ga_2O_3$ SBDs fabricated with (black) and without (blue) exposure to the 1050 °C LPCVD thermal cycle used for the Ga-assisted etching process. (a) Forward and reverse J-V characteristics. (b) Reverse breakdown characteristics. (c) C-V and $1/C^2$-V characteristics. (d) Extracted net donor concentration profiles.

Figure 5 shows the temperature-dependent transport of the planar and LPCVD mesa-terminated diodes from 25 to 250 °C. In semilog scale [Fig. 5(a)], both structures retain strong rectification at all temperatures, with forward current density increasing with temperature in the low-to-moderate bias range-behavior consistent with thermionic-emission (TE) transport across the Ni/β-$Ga_2O_3$ barrier [63, 69]. The nearly parallel forward characteristics on linear scale [Fig. 5(b)] indicate comparable series resistance in planar and mesa devices over the full temperature range. The evolution of the specific on-resistance, $R_{on,sp}$, extracted from the linear forward regime is shown in Figs. 5(c)-(d). For both devices, $R_{on,sp}$ exhibits the well-known competition between injection and drift-limited transport: an initial reduction at modest temperature and low bias due to enhanced TE injection and contact conductivity, followed by a monotonic increase with temperature once drift resistance dominates as electron-phonon scattering lowers the β-$Ga_2O_3$ mobility [61, 72, 73]. The mesa-etched diodes track the planar reference closely at each temperature, with only a small offset that mirrors the slight $V_{turn_on}$ shift discussed earlier. The net increase in $R_{on,sp}$ from 25 to 250 °C is comparable to prior measurements on β-$Ga_2O_3$ SBDs, where phonon-limited mobility degradation dominates at high temperatures [61, 72, 73]. On the reverse side of Fig. 5(a), the leakage current increases modestly with temperature, as expected for TE and Thermionic-Field Emission (TFE) components in Schottky junctions. Across the full temperature range, the mesa-terminated devices show lower leakage than the planar controls. This behavior is consistent with the marginally higher SBH extracted for the mesa structure from both J-V and C-V measurements, which would be expected to reduce reverse current. Importantly, even at 250 °C, the mesa-terminated diodes maintain stable forward and reverse characteristics without leakage escalation or barrier instability, indicating that the LPCVD deep-etch process preserves the junction and drift-layer transport properties.

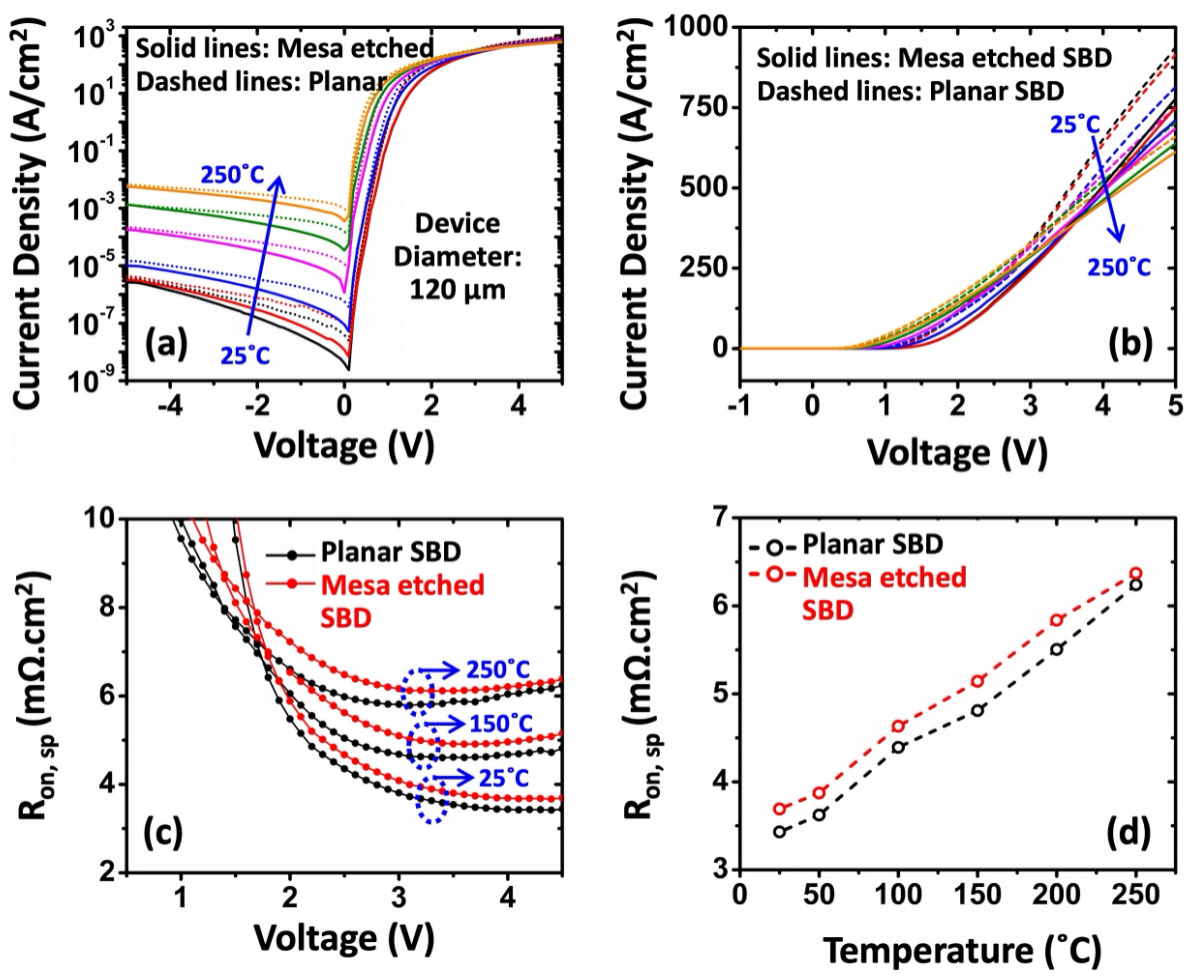

Fig. 5. Temperature-dependent J-V characteristics of planar and mesa-etched β-$Ga_2O_3$ SBDs in (a) log scale and (b) linear-scale. (c) Extracted specific on-resistance versus voltage at different temperatures. (d) Temperature dependence of specific on-resistance, showing similar trends for planar and mesa-etched diodes.

Figure 6(a) shows the temperature-dependent Schottky barrier height ($\Phi_B$) and ideality factor (η) extracted from forward J-V characteristics for the planar and mesa-etched diodes. For both structures, $\Phi_B$ exhibits a negative temperature dependence, decreasing from 1.15 eV to 1.11 eV for the planar device and from 1.16 eV to 1.12 eV for the mesa-etched device between 25 °C and 250 °C. This trend is consistent with previous reports on β-$Ga_2O_3$ Schottky contacts [63, 67], where spatial variations at the metal-semiconductor interface lead to a gradual decrease in the extracted $\Phi_B$ at elevated temperature. Across the full temperature range, the mesa-terminated diode maintains a slightly higher $\Phi_B$ than the planar reference, typically by a few meV, in agreement with the higher C-V extracted barrier height (Fig. 3) and the lower reverse leakage observed in Figs. 2 and 5. The ideality factor increases monotonically with temperature for both devices, rising from 1.20 to 1.33 for the planar diode and from 1.21 to 1.33 for the mesa device between 25 °C and 250 °C. This gradual increase in η is also consistent with thermionic emission in the presence of a spatially non-uniform barrier, where thermally activated transport through regions of varying local $\Phi_B$ and possible recombination-assisted components leads to a deviation from ideal behavior at elevated temperature [67]. The close agreement between the planar and mesa curves for both $\Phi_B(T)$ and η(T) indicates that the deep etch in LPCVD system does not appear to introduce measurable interface perturbations, and the temperature dependence of the Schottky contact is influenced primarily by the intrinsic barrier inhomogeneity of the Ni/β-$Ga_2O_3$ interface.

Figure 6(b) shows the temperature dependence of the reverse leakage current density and the corresponding on/off current ratio for the planar and mesa-etched diodes. As temperature increases, both devices exhibit the expected rise in reverse current. The mesa-etched diode consistently shows reduced leakage compared to the planar structure; for example, at -5 V and 25 °C, the leakage in the mesa device is roughly two-thirds of that of the planar diode, and a similar reduction is observed at 100 °C, where the planar device reaches the mid-$10^{-5}$ A/cm² range while the mesa current remains lower. At the highest measurement temperature of 250 °C, the mesa diode consistently exhibits a modest reduction in leakage relative to the planar device, which aligns with its slightly higher extracted SBH. The on/off ratio follows the inverse trend, decreasing with temperature as reverse leakage increases. At 25 °C, both devices exhibit rectification ratios on the order of $10^8$, reflecting good room-temperature Schottky behavior. Although the rectification ratio decreases with increasing temperature, both structures maintain values above $10^5$ at 250 °C, demonstrating stable high-temperature operation, with the mesa diode showing slightly higher rectification due to its lower leakage. The ability of both devices to retain strong rectification across the entire 25-250 °C range indicates that the plasma-free LPCVD mesa formation preserves the high-temperature reliability of the Schottky junction.

To evaluate the effectiveness of the LPCVD-defined mesa as a field-termination structure, reverse-bias breakdown measurements were performed on both planar and mesa-terminated β-$Ga_2O_3$ Schottky diodes as shown in Figure 7(a). The planar device undergoes breakdown at 287 V, which is consistent for this drift-layer design in the absence of any field-termination technique [73, 74]. In contrast, the mesa-terminated diode exhibits an increased breakdown voltage of 500 V under reverse bias. This enhancement confirms that deep LPCVD mesa etching effectively redistributes the electric field at the anode perimeter, mitigating edge crowding that typically limits the breakdown performance of planar vertical SBDs. Using the one-dimensional electrostatic estimate for the electric field [75], the planar device at a donor concentration of $2 \times 10^{16}$ $cm^{-3}$ corresponds to a parallel plate breakdown field strength of 1.44 $MV \cdot cm^{-1}$, while the mesa-terminated device yields 1.90 $MV \cdot cm^{-1}$, indicating a ~32% increase in breakdown electric field enabled by geometric field redistribution from the LPCVD-defined mesa termination. Further improvements are expected through optimization of mesa geometry and integration with field-management strategies such as field plates or junction termination extension [76].

The underlying mechanism for the enhanced blocking capability of the mesa-terminated diode is investigated through the two-dimensional electric-field simulations shown in Figures 7(b)-7(d). For the planar structure, the simulated reverse-bias field distribution reveals a pronounced peak located at the perimeter of the Schottky anode. This peak arises from the convergence of lateral and vertical depletion regions and results in a highly nonuniform field profile, with the maximum electric field concentrated within a narrow region adjacent to the metal edge. Such edge-dominated field crowding has been widely recognized as the primary factor limiting breakdown voltage in β-$Ga_2O_3$ vertical Schottky diodes [76], particularly when no termination scheme is implemented. When the deep LPCVD-etched mesa is introduced, the electrostatic landscape changes substantially. The recessed geometry shifts the depletion boundary below the original surface contour, allowing the electric field to spread more uniformly into the bulk of the drift region rather than concentrating at the surface. As shown in Fig. 7(b), the high-field zone that dominates the planar device is markedly reduced at the anode edge in the mesa-terminated structure, with the peak field relocated deeper inside the drift layer. The reduction in surface field intensity directly correlates with the experimentally observed ability of the mesa-terminated device to sustain reverse biases up to 500 V before breakdown. The dependence of field suppression on mesa depth, presented in the vertical profile analysis of Fig. 7(c), also confirms this interpretation. For shallow etch depths, the simulated field at the Schottky edge remains high, and an additional peak appears at the mesa corner. As the mesa depth increases, both the edge field and the corner field decrease rapidly. Depths exceeding ~ 4-6 μm effectively suppress these localized maxima, while depths approaching 8 μm, consistent with the experimental structure, provide substantial field relief with diminishing improvement at greater depths.

To assess the influence of mesa-to-anode alignment, additional simulations were performed for different mesa-to-anode edge separations, as shown in Fig. 7(d). The results reveal a strong dependence of the peak electric field on the separation distance. The edge-terminated structure exhibits the lowest peak electric field of 3.19 MV/cm. As the mesa-to-anode separation increases, the peak electric field rises rapidly to 6.18 MV/cm at 0.1 μm, 7.26 MV/cm at 0.2 μm, and 9.45 MV/cm at 1 μm, eventually approaching the planar-device value of 9.88 MV/cm for larger separations. These results indicate that even small mesa-to-anode separations can substantially influence the electric-field distribution and field-termination efficiency. Since the Ga-assisted etching process is crystallographically anisotropic, local variations in the mesa edge profile may be present around the device perimeter, resulting in non-uniform mesa-to-anode spacing. The simulation results therefore provide insight into the sensitivity of the field distribution to local geometric variations. Further improvements are expected through optimization of the etching process and mesa geometry, as well as by fabricating devices on crystallographic orientations that exhibit reduced lateral etching and more uniform etch profiles, leading to improved field distribution. Despite these non-idealities, the experimentally observed increase in breakdown voltage, without significant degradation in forward conduction or leakage characteristics, indicates that the overall field redistribution provided by the mesa structure remains the dominant effect, highlighting the effectiveness of LPCVD-based, plasma-free mesa termination for improving the blocking performance of β-$Ga_2O_3$ power devices.

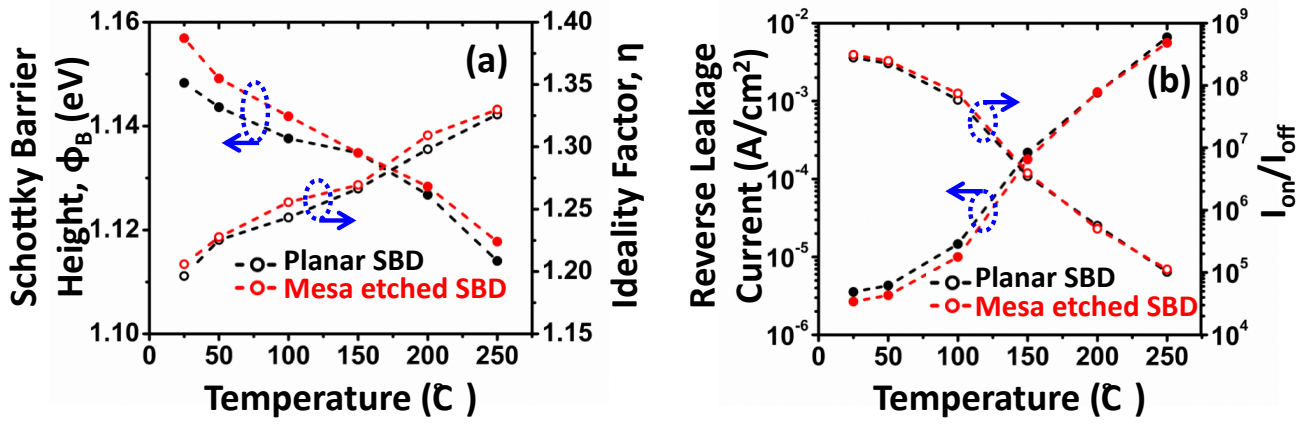

Fig. 6. (a) Temperature-dependent Schottky barrier height and ideality factor for planar and mesa-etched β-$Ga_2O_3$ SBDs. (b) Temperature dependence of reverse leakage current and corresponding on/off ratio.

## IV. CONCLUSION

In summary, the results demonstrate that Ga-assisted LPCVD deep etching provides an effective, plasma-free approach for implementing field termination in vertical β-$Ga_2O_3$ SBDs. The LPCVD process, driven purely by a thermochemical sub-oxide reaction, eliminates plasma-induced sidewall damage while maintaining excellent diode performance. Electrical characterization revealed that the forward conduction of the mesa-terminated devices remains nearly identical to that of planar controls, confirming that the LPCVD process preserves the integrity of the Schottky interface. The specific on-resistance remained stable, validating that carrier transport in the drift layer was unaffected by the deep-etch process. Capacitance-voltage analysis confirmed a uniform donor concentration of $2 \times 10^{16}$ $cm^{-3}$ across both device types and no signs of depletion broadening or charge compensation. Temperature-dependent transport characteristics further indicate the robustness of the LPCVD-processed diodes. The current density increases with temperature following thermionic-emission behavior, while the on-resistance exhibits the characteristic transition from injection-limited conduction at lower temperatures to phonon-limited drift transport at higher temperatures. The barrier height gradually decreases from 1.16 to 1.12 eV, and the ideality factor shows a rise from 1.21 to 1.33, behavior consistent with transport across spatially inhomogeneous yet stable Schottky interfaces. Leakage current remained low and rectification ratios exceeded $10^5$ up to 250 °C. An improvement was observed under reverse bias, where the LPCVD-terminated diodes achieved breakdown voltage of 500 V compared to 287 V for planar devices. The increase in blocking capability directly reflects the redistribution of the electric field at the Schottky edge, verified through TCAD simulations that revealed a substantial reduction in peak field intensity and a smoother potential gradient across the drift region. Overall, these findings demonstrate that the Ga-assisted LPCVD deep etching can effectively modify the device geometry to suppress edge-field crowding while preserving the intrinsic electronic quality of β-$Ga_2O_3$. The electrical stability across temperature, the maintenance of low on-resistance, and the enhanced breakdown voltage together establish LPCVD deep etching as a reliable, thermally robust, and scalable field-termination approach for future high-voltage β-$Ga_2O_3$ power devices.

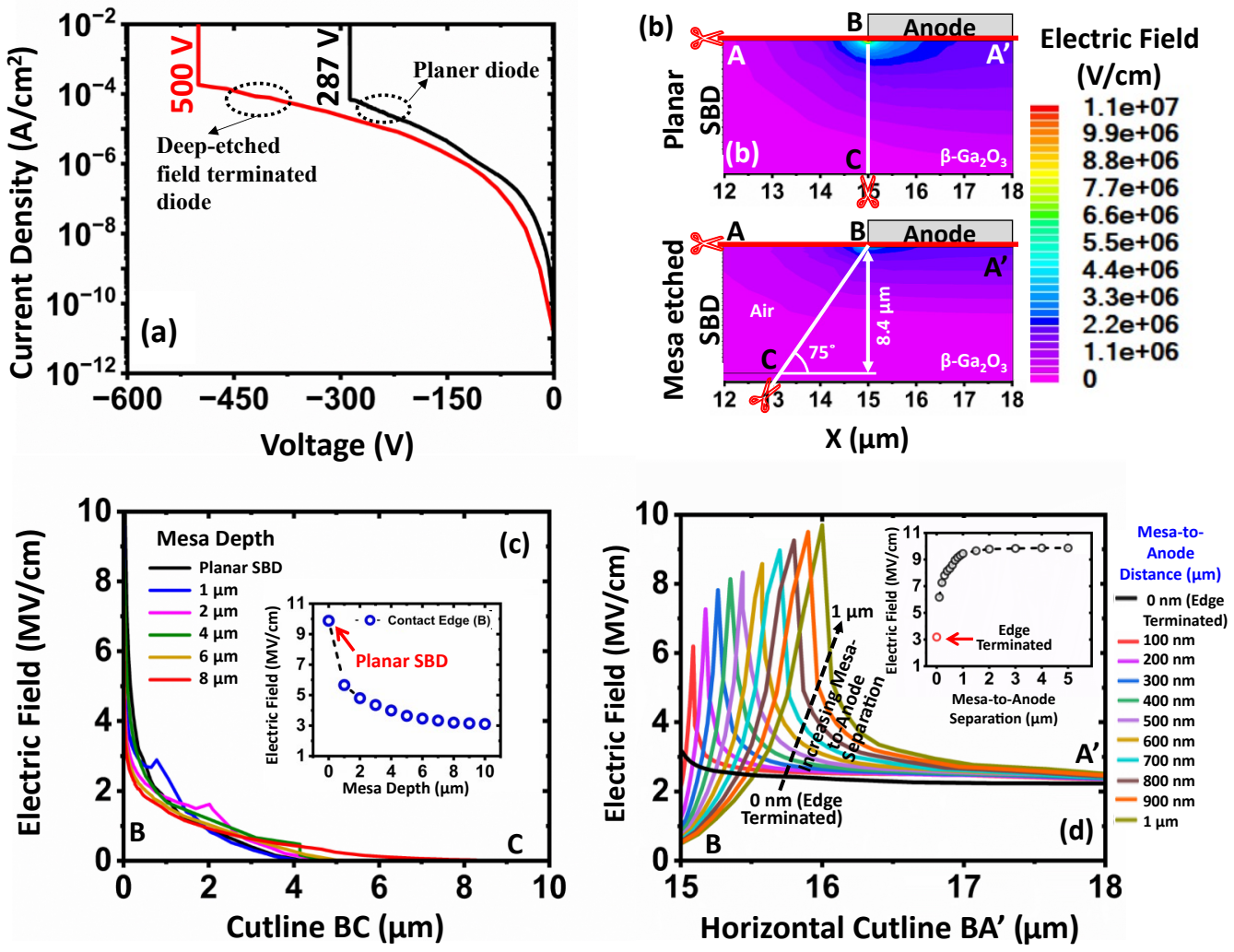

Fig.7. (a) Reverse J-V characteristics of planar and deep mesa-terminated β-$Ga_2O_3$ SBDs. Insets: (b) Simulated two-dimensional electric-field distributions for planar and mesa-terminated structures under 500 V reverse bias. (c) Simulated electric-field profiles along cutline B-C for different mesa depths. Inset: Peak electric field at the contact edge as a function of mesa depth. (d) Simulated electric-field profiles along cutline B-A′ for different mesa-to-anode edge separations. Inset: Peak electric field as a function of mesa-to-anode separation.

## Acknowledgments

This work was supported by the National Science Foundation under Award Nos. ECCS-2501623 and ECCS-2532898.The authors acknowledge Jiawei Liu from Professor Uttam Singisetti's research group at University at Buffalo for assistance with the breakdown measurements.